\documentclass[
    ,final            
  ]
  {aipproc}

\layoutstyle{6x9}
\usepackage{subfigure}

\usepackage{floatflt}
\usepackage{wrapfig}

\begin{document}

\title{AGN Feedback and Bimodality in Cluster Core Entropy}


\author{Fulai Guo}{
  address={Dept of Physics, University of California, Santa Barbara, CA 93106}
}

\author{S. Peng Oh}{
  address={Dept of Physics, University of California, Santa Barbara, CA 93106}
}

\author{M. Ruszkowski}{
  address={Dept of Astronomy, University of Michigan, Ann Arbor, MI 48109}
}

\begin{abstract}
We investigate a series of steady-state models of galaxy clusters, in which the hot intracluster gas is efficiently heated by active galactic nucleus (AGN) feedback and thermal conduction, and in which the mass accretion rates are highly reduced compared to those predicted by the standard cooling flow models. We perform a global Lagrangian stability analysis. We show for the first time that the global radial instability in cool core clusters can be suppressed by the AGN feedback mechanism, provided that the feedback efficiency exceeds a critical lower limit. 
Furthermore, our analysis naturally shows that the clusters can exist in two distinct forms.
Globally stable clusters are expected to have either: 1) cool cores 
stabilized by both AGN feedback and conduction, or 2) non-cool cores stabilized primarily by conduction. Intermediate central temperatures typically lead to globally unstable solutions. 
This bimodality is consistent with the recently observed anticorrelation between the flatness of the temperature profiles and the AGN activity \citep{dunn08} 
and the observation by Rafferty et al. (2008) \citep{rafferty08} 
that the shorter central cooling times tend to correspond to 
significantly younger AGN X-ray cavities.
\end{abstract}

\maketitle


\section{Introduction}

Recent high-resolution Chandra and XMM-Newton observations of cluster cool cores show a remarkable lack of emission lines from the gas at temperatures below about ~ 1/3 of the ambient cluster temperature, which is suggestive of one or more heating mechanisms maintaining the hot gas at keV temperatures for a period at least comparable to the lifetime of galaxy clusters. Since global thermal instability may result in a cooling catastrophe and a strong cooling flow, a successful quasi-equilibrium model for the ICM must be globally stable, or at least only have instabilities which grow on extremely long timescales. In \citet{Guo08b}, we performed a formal global Lagrangian stability analysis (in the spirit of that performed by \citep{kim03}) to investigate thermal instability in quasi-equilibrium galaxy clusters with thermal conduction and AGN feedback heating. 

\section{Methods}

In our model, the ICM is subject to radiative cooling, AGN feedback heating and thermal conduction. For thermal conduction, we adopt Spitzer conductivity with a factor f due to magnetic field suppression; for AGN heating, we adopt the effervescent heating mechanism proposed by \citep{begelman01,ruszkowski02} and assume that the AGN mechanical luminosity is proportional to the central mass accretion rate: $L_{\rm AGN}= \epsilon \dot{M}_{\rm in} c^{2}$, where $\epsilon$ is the kinetic efficiency of AGN accretion. 

To build initial background states for stability analysis, we first construct steady-state cluster models, where the gas density and temperature profiles fit observations quite well. We take the cluster Abell 1795 as our fiducial cluster. The  gas cooling time is much less than the age of the cluster within $\sim$100 kpc from the cluster center.

We linearize the gas hydrodynamic equations by using the Lagrangian perturbation method.
Here we only consider radial perturbations. A key ingredient of our model is that the AGN heating rate is proportional to the central mass accretion rate. The perturbation of the AGN heating rate due to the feedback mechanism in the energy perturbation equation is found to play 
a key role in suppressing global thermal instability in cool core clusters. The perturbed hydrodynamic equations form an eigenvalue problem, which can be solved numerically with appropriate boundary conditions to find global eigenmodes and the growth rate (which is an eigenvalue of the problem). The detailed equations and numerical methods are shown in our paper, which also performs a local analysis of small-scale modes.

\section{Results}

\begin{figure}[ht]
\resizebox{!}{13.0cm}{\includegraphics{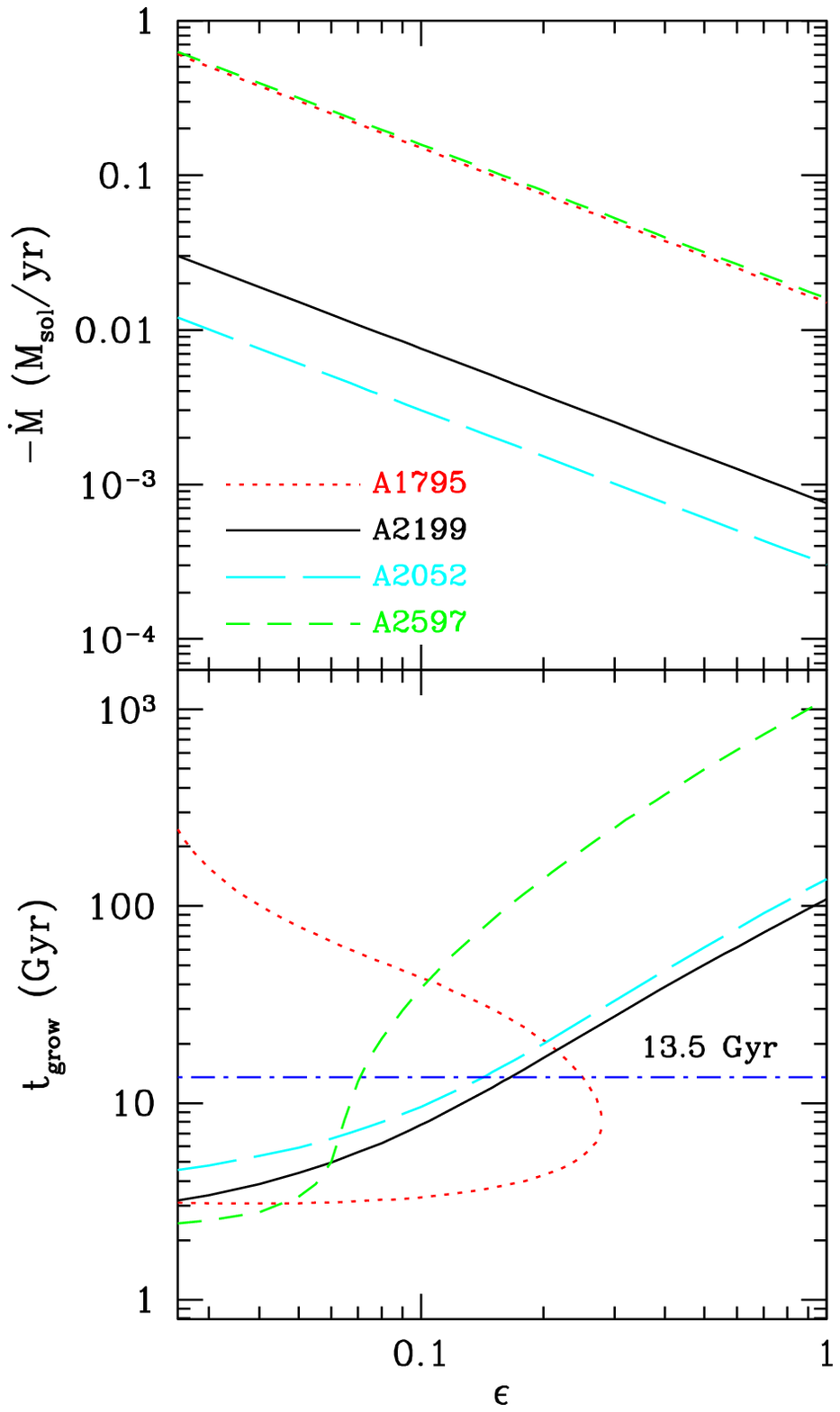}}
\resizebox{!}{13.0cm}{\includegraphics{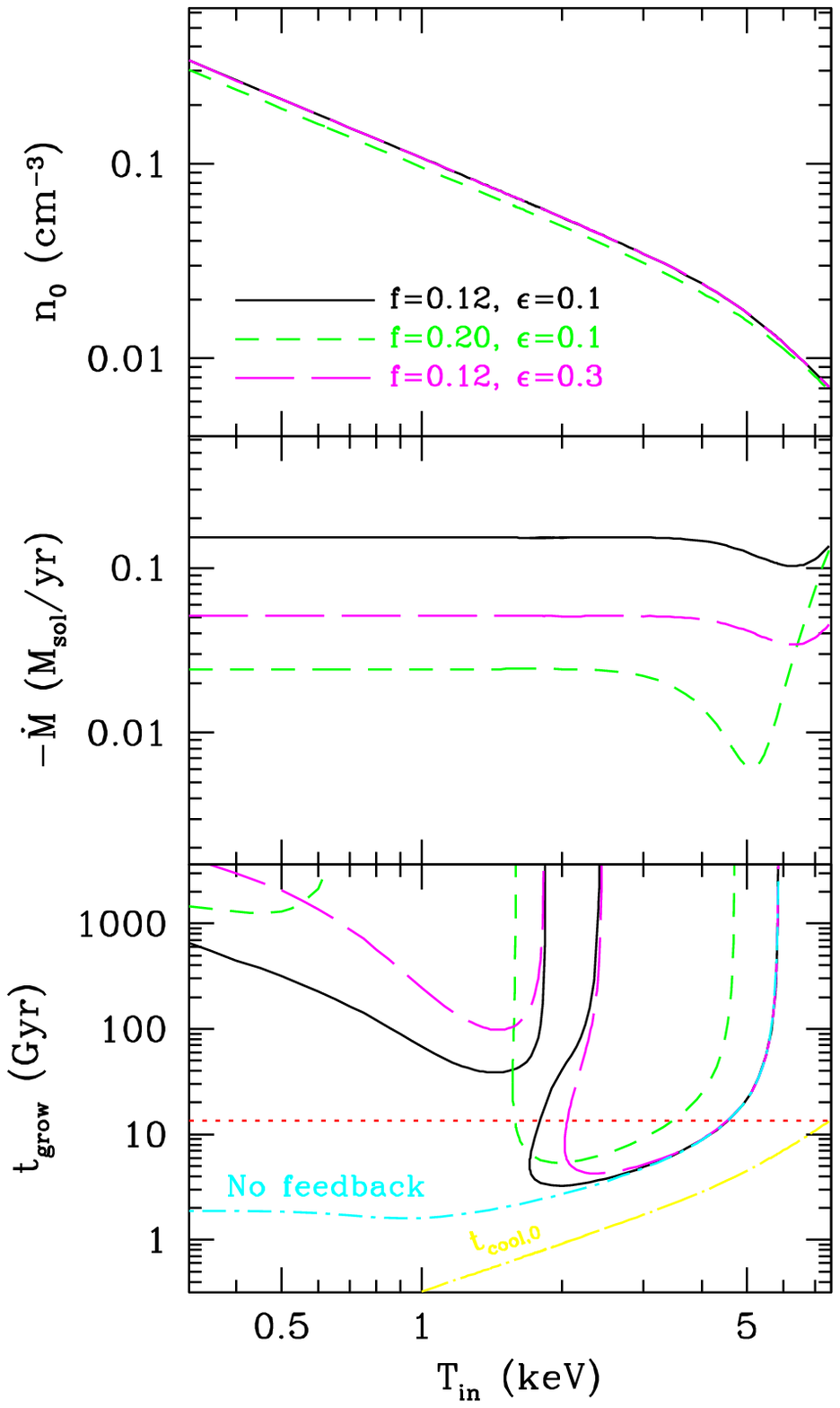}}
\caption{{\bf Left panel}: Effect of the AGN feedback efficiency on thermal stability in typical cool core clusters. For different models of each cluster, the values of $f$ and $\epsilon \dot{M}$ are roughly the same (see paper for details). Top panel: scaling of $\dot{M}$ with $\epsilon$. Bottom panel: the dependence of the growth time of unstable modes on $\epsilon$ for each cluster. Note that the clusters are stable or effectively stable when $\epsilon$ is greater than a lower limit $\epsilon_{\rm min}$. The line for A1795 is double-valued since it has two unstable modes; both these modes vanish when $\epsilon > 0.28$. Here, we assume $\epsilon$ is a constant; if, as suggested by observations, $\epsilon \propto \dot{M}^{\nu}$ (with $\nu \sim 0.3-0.6$), then $\epsilon_{\rm min}$ will be smaller. {\bf Right panel}: Dependence of the cluster stability on the background profile for the cluster A1795. For a fixed value of $f$ and $\epsilon$, the corresponding central electron number density (\textit{upper panel}), steady-state mass accretion rate (\textit{middle panel}), and the growth time of unstable modes (\textit{lower panel}) are plotted as a function of the central gas temperature $T_{\rm{in}}$. The \textit{dot short-dashed} line in the lower panel shows the growth time of the unstable mode in steady-state models with $f=0.12$, $\epsilon=0.1$ and no feedback mechanism for AGN heating (see paper for details), while the \textit{dot long-dashed} line stands for the central gas cooling time in models with $f=0.12$, $\epsilon=0.1$.}
\label{fig:main_results} 
\end{figure}

One might conjecture that the ICM is stable (or effectively stable) if the AGN feedback efficiency is greater than a lower limit. We thus explore the dependence of the cluster stability on the AGN feedback efficiency, which is shown in the left panel of Fig. \ref{fig:main_results}. The lower left panel of Fig. \ref{fig:main_results} clearly shows that the cluster is stable or effectively stable  when $\epsilon$ is greater than a lower limit $\epsilon_{\rm min}$. The values of $\epsilon_{\rm min}$ for these four typical cool core clusters are $\epsilon_{\rm min} \sim 0.1-0.3$, which is roughly consistent with the recent estimate of $\epsilon\sim 0.3$ by \citep{heinz07} and is marginally consistent with observational estimates of $\epsilon \sim 0.01-0.1$ by \citep{allen06,merloni07}. 

We study the dependence of the cluster stability on the background steady-state ICM profiles. We consider the models of the cluster A1795 with f = 0.12 and $\epsilon$ = 0.1. 
The right column of Fig. 1 shows three typical steady-state cluster profiles. 
The lower right panel of Fig. 1 shows the dependence of the cluster stability on the background profile for A1795. Obviously, either non-cool core clusters with relatively flat temperature profiles or cool core clusters with relatively steep temperature profiles are stable (or effectively stable), while either a conductive or thermal instability will develop in clusters with intermediate temperature profiles, driving them to either extremes. 

The dot-dashed lines in the lower right panel of Figure 1 show the growth time of the unstable mode in steady-state models without the feedback mechanism for AGN heating. 
In this case, the instability growth time in cool core clusters with relatively steep temperature profiles is very short ($\sim$2 Gyr), suggesting that the feedback mechanism plays a key role in stabilizing thermal instability in these clusters. On the other hand, although the stabilizing effect of the feedback mechanism becomes negligible for non-cool core clusters with relatively flat temperature profiles, our stability analysis surprisingly shows that these cluster models are also stable. Even ignoring the AGN heating in the perturbed energy equation, we usually still find that these non-CC clusters are stable, which suggests that thermal conduction alone completely suppresses thermal instability in these clusters. Thus, thermal stability of the ICM favors two distinct categories of cluster steady state profiles: cool core clusters stabilized mainly by AGN feedback and non-CC clusters stabilized by thermal conduction. Interestingly, X-ray observations also suggest that clusters can be subdivided into two distinct categories according to the presence or absence of a cool core (see \citep{sanderson06} and references therein). Our stability analysis naturally explains these two distinct cluster categories. 

Subsequently, we also investigated whether AGN outbursts could transform cool core clusters in \citet{guo09}. We consider if time-variable conduction and AGN outbursts could be responsible for the cool-core (CC), non cool-core (NCC) dichotomy. We show that strong AGN heating can bring a CC cluster to a NCC state, which can be stably maintained by conductive heating from the cluster outskirts. On the other hand, if conduction is shut off by the heat-flux driven buoyancy instability, then the cluster will cool to the CC state again, where it is stabilized by low-level AGN heating. Thus, the cluster cycles between CC and NCC states. In contrast with massive clusters, we predict the CC/NCC bimodality should vanish in groups, due to the lesser role of conductive heating there. We find tentative support from the distribution of central entropy in groups, though firm conclusions require a larger sample carefully controlled for selection effects. Please refer to the paper \citep{guo09} for more details.


\bibliographystyle{aipproc}   

\bibliography{ms}

\end{document}